\documentclass[final,5p,times,twocolumn,preprint]{elsarticle}
\usepackage{slashed}
\usepackage{eucal} 
\usepackage{bm}
\usepackage[hypertexnames=false, colorlinks=true, citecolor=blue, urlcolor=blue, linkcolor=black]{hyperref}
\usepackage{amsmath,amssymb,amsfonts,latexsym}
\newcommand{\Ithreep}{{I'}_{\hspace{-.6ex}3}}
\newcommand{\Sthree}{S_{\hspace{-.3ex}3}}
\newcommand{\Sthreep}{{S'}_{\hspace{-.8ex}3}}
\allowdisplaybreaks
\begin{document}
\begin{frontmatter}
\title{QCD angular momentum in $N \rightarrow \Delta$ transitions}
\author[j]{June-Young Kim}
\ead{jykim@jlab.org}
\author[i]{Ho-Yeon Won}
\ead{hoywon@inha.edu}
\author[h,j]{Jose~L.~Goity}
\ead{goity@jlab.org}
\author[j]{Christian~Weiss}
\ead{weiss@jlab.org}
\address[j]{Theory Center, Jefferson Lab, Newport News, VA 23606, USA}
\address[i]{Department of Physics, Inha University, Incheon 402-751, South Korea} 
\address[h]{Department of Physics, Hampton University, Hampton, VA 23668, USA}
\begin{abstract}
$N \rightarrow \Delta$ transitions offer new possibilities for exploring the isovector component of the
QCD quark angular momentum (AM) operator causing the $J^{u - d}$ flavor asymmetry in the nucleon.
We extend the concept of QCD AM to transitions between baryon states, using light-front densities
of the energy-momentum tensor in transversely localized states.
We calculate the $N \rightarrow \Delta$ transition AM in the $1/N_c$ expansion, connect it with the
$J^{u - d}$ flavor asymmetry in the nucleon, and estimate the values using lattice QCD results.
In the same setup we connect the transition AM to the transition GPDs sampled in hard exclusive
electroproduction processes with $N \rightarrow \Delta$ transitions, enabling experimental study
of the transition AM.
\end{abstract}
\end{frontmatter}
\section{Introduction}
Angular momentum (AM) has become an essential concept in hadron structure physics.
The AM operator is derived from the QCD energy-momentum tensor (EMT) and represents the conserved
current associated with rotational invariance. It measures the AM of chromodynamic field configurations,
arising from their space-time dependence (orbital AM) and internal degrees of freedom (spin), and can
be decomposed into quark and gluon contributions. Its formal properties have been discussed extensively
and are now well understood; see Refs.~\cite{Leader:2013jra,Lorce:2017wkb} for reviews.
Its experimental study becomes possible through the connection with the generalized parton distributions 
(GPDs) describing hadron structure as probed in high-momentum-transfer exclusive scattering processes;
see Refs.~\cite{Goeke:2001tz,Diehl:2003ny,Belitsky:2005qn,Kumericki:2016ehc} for reviews.
Certain components of the EMT can be expressed as integrals of the GPDs (moments) and thus indirectly
be related to observables measured in exclusive processes \cite{Ji:1996ek,Polyakov:2002yz}.

There is evidence of a large flavor asymmetry of the quark AM in the nucleon, $J^{u - d}$.
The normalization of the Pauli form factor-type GPD $E$ entering in the AM
sum rules \cite{Ji:1996ek,Polyakov:2002yz} is controlled by the nucleon anomalous magnetic moment,
whose isovector component is much larger than the isoscalar, $\kappa^{p - n} = 3.7$ vs.\
$\kappa^{p + n} = -0.12$. Lattice QCD calculations of the quark AM show large
flavor asymmetries \cite{Gockeler:2003jfa,LHPC:2007blg,LHPC:2010jcs,Bali:2018zgl,Alexandrou:2019ali}.
GPD models consistent with present experimental
data also suggest a large flavor asymmetry \cite{Goeke:2001tz,Diehl:2003ny,Belitsky:2005qn,Kumericki:2016ehc}.
The question of ``isovector AM'' is central to the understanding of nucleon structure and
nonperturbative dynamics and needs further study.

Like any local composite operator in QCD, the quark flavor components of the EMT have matrix elements
not only between states of the same hadron (form factors) but also between states of different hadrons
(transition form factors).
This makes it possible to formulate a concept of AM for transitions between hadronic states.
Of particular interest is the AM in $N \rightarrow \Delta$ transitions. Because the isospin difference
is $\Delta I = 1$, the transition AM is a pure isovector and thus related to the $J^{u - d}$
flavor asymmetry in the nucleon. Because the structure of the $N$ and $\Delta$ baryons
is closely connected, study of the transition AM can provide further insight into nucleon structure.
The transition AM can be connected with the GPDs sampled in hard exclusive processes with
$N \rightarrow \Delta$ transitions, enabling its experimental study
\cite{Semenov-Tian-Shansky:2023bsy,Kroll:2022roq,CLAS:2023akb}.

The $1/N_c$ expansion of QCD is a powerful method for analyzing the spin-flavor structure
of hadronic matrix elements of QCD operators such as the EMT and AM \cite{tHooft:1973alw,Witten:1979kh}.
It establishes a hierarchy among the spin-flavor components of the $N \rightarrow N$ matrix elements of the EMT.
It also connects the $N \rightarrow N$ and $N \rightarrow \Delta$ (and even $\Delta \rightarrow \Delta$)
matrix elements of the EMT through the emergent spin-flavor symmetry in large-$N_c$ limit
\cite{Gervais:1983wq,Gervais:1984rc,Dashen:1993as,Dashen:1993jt,Dashen:1994qi}.
The method is therefore uniquely suited for analyzing the flavor structure of QCD AM in the
nucleon and exploring its extension to $N \rightarrow \Delta$ transitions \cite{Goeke:2001tz}.

In this letter we study the isovector QCD AM in $N \rightarrow \Delta$ transitions and its connection with
the $J^{u - d}$ flavor asymmetry in the nucleon. We formulate the concept of transition AM using light-front
densities of the EMT in transversely localized baryon states. We calculate the $N \rightarrow \Delta$ transition
AM in the $1/N_c$ expansion, connect it with the $J^{u - d}$ flavor asymmetry in the nucleon, and estimate
its numerical value using lattice QCD results. In the same setup we connect the transition AM to the
GPDs sampled in hard exclusive processes with $N \rightarrow \Delta$ transitions.
\section{Transition angular momentum}
\label{sec:transition_am}
The definition of the QCD AM operator and the interpretation of its $N \rightarrow N$ matrix elements
have been discussed extensively in the literature \cite{Leader:2013jra,Lorce:2017wkb}.
The extension to transitions $B \rightarrow B'$ between baryon states with different mass and spin
raises new questions that require fresh consideration. The definition of the nucleon AM of
Ref.~\cite{Ji:1996ek} uses the specific form of the $N \rightarrow N$ matrix element of the EMT
and cannot immediately be extended to $B \rightarrow B'$ transitions. The definition of the nucleon
AM density of Ref.~\cite{Polyakov:2002yz} uses the Breit frame and assumes heavy nucleons
(non-relativistic motion) and becomes ambiguous for transitions between states with different mass.

The formulation of AM as a transverse density at fixed light-front time \cite{Lorce:2017wkb}
offers a natural framework for the extension to $B \rightarrow B'$ transitions.
The light-front formulation is fully relativistic and can be extended to transitions between states
with different mass and spin. It permits the preparation of transversely localized states independently
of their mass, using the effectively non-relativistic kinematics in the transverse space.
It contains a prescription for defining the hadron spin states through the light-front helicity,
which enables consistent spin decomposition of the matrix elements.
It also provides a simple mechanical picture of the longitudinal AM density as the cross product
of transverse position and the momentum density measured by the EMT, which can be applied directly
to $B \rightarrow B'$ transitions (see below). Transverse densities for $N \rightarrow \Delta$ transitions
have been used successfully in the description of electromagnetic structure \cite{Carlson:2007xd}.
Here we employ this formulation to define the AM in $B \rightarrow B'$ transitions and discuss its properties.

In the following we use the symmetric (Belinfante-improved) version of the EMT, which gives rise to an
AM operator measuring the total AM; the separation of spin and orbital AM is discussed
below \cite{Leader:2013jra,Lorce:2017wkb}. The operator describing the contribution of quarks
with flavor $f$ is
\begin{align}
\hat{T}^{\alpha\beta}_f (x) &=
i \bar\psi_f (x) \gamma^{\{\alpha} \overleftrightarrow{\nabla}^{\beta\}} \psi_f (x),
\label{T_symmetric}
\end{align}
where $\overleftrightarrow{\nabla}^{\mu} \equiv \frac{1}{2} (\overrightarrow{\partial}^{\mu}
- \overleftarrow{\partial}^{\mu}) - igA^{\mu}$ is the covariant derivative and $\{\alpha\beta \} \equiv
\frac{1}{2} (\alpha\beta + \beta\alpha)$;
the operator for gluons is given in Ref.\cite{Lorce:2017wkb} and not needed here. We assume two quark flavors
and define the isoscalar and isovector components as
\begin{align}
(\hat{T}^{V, S})^{\alpha\beta} \; &\equiv \;
\hat{T}_u^{\alpha\beta} \; \pm \;
{{\hat{T}_d}\hspace{-1em}\phantom{T}^{\alpha\beta}} .
\label{isoscalar_isovector}
\end{align}
Note that these quark operators are not conserved currents; only the sum of the isoscalar quark and gluon EMT
is a conserved current obtained from Noether's theorem.
We consider the transition matrix elements of the operators Eq.~(\ref{isoscalar_isovector})
between general baryon states with masses $m$ and $m'$ and 4-momenta $p$ and $p'$,
\begin{align}
\langle B', p' | \hat{T}^{\alpha\beta}(0) |B, p \rangle ,
\label{T_matrix}
\end{align}
where $B \equiv \{S, \Sthree, I, I_3\}$ and $B' \equiv \{S',\Sthreep,I',\Ithreep\}$ collectively denote the
spin-isospin quantum numbers. The choice of spin states and the spin-isospin dependence of the
matrix element are discussed below. The 4-momentum transfer is $\Delta \equiv p' - p$, the invariant momentum
transfer $t \equiv \Delta^2$, and the average baryon 4-momentum is $P \equiv (p' + p)/2$.
The 4-vectors and tensors are described by the light-front
components $p^\pm \equiv p^0 \pm p^3, \bm{p}_T \equiv (p^1, p^2)$.
We consider Eq.~(\ref{T_matrix}) in a class of frames where $\Delta^+ = 0$ and $\bm{P}_T = 0$
(generalized Drell-Yan-West frame). In these frames
$t = - \bm{\Delta}_T^2 < 0$.\footnote{In the matrix element
Eq.~(\ref{T_matrix}) the invariant momentum transfer can also attain timelike values $0 < \Delta^2 < (m' - m)^2$.
The following definition of the transition AM and its density refers to the spacelike part of the form factors.
This is consistent with the standard definition of the transition magnetic moment through the $t = 0$ magnetic
transition form factor in electromagnetic processes \cite{Jones:1972ky}.
In the $1/N_c$ expansion of $N$--$\Delta$ transition
matrix elements, $(m_\Delta - m_N)^2 = \mathcal{O}(N_c^{-2})$ is strongly suppressed.}
In the notation $p = [p^+, p^-, \bm{p}_T]$, the momentum components are given by
\begin{align}
p = \left[ p^+, \frac{m^2 + |\bm{\Delta}_T|^2/4}{p^+}, -\frac{\bm{\Delta}_T}{2} \right] ,
\nonumber \\
p' = \left[ p^+, \frac{m^{\prime 2} + |\bm{\Delta}_T|^2/4}{p^+}, \frac{\bm{\Delta}_T}{2} \right] ,
\nonumber \\
\Delta = \left[ 0, \frac{m^{\prime 2} - m^2}{p^+}, \bm{\Delta}_T \right] .
\label{frame_lf}
\end{align}
$p^+$ remains undetermined, and its choice selects a particular frame in the class (boost parameter).
The matrix element Eq.(\ref{T_matrix}) becomes function of $\bm{\Delta}_T$. For constructing the AM, we take
the $+i$ ($i$ = 1, 2) component of the EMT 
\begin{align}
T^{+i}(\bm{\Delta}_T| B', B) & \equiv \langle B', p' | \hat{T}^{+i}(0) |B, p \rangle ,
\label{T_matrix_Delta}
\end{align}
and define a transverse coordinate density as
\begin{align}
T^{+i}(\bm{b}| B', B)
\equiv \int \frac{d^2 \Delta_T}{(2\pi )^2} e^{-i\bm{\Delta}_T \bm{b}} T^{+i}(\bm{\Delta}_T| B', B) .
\end{align}
This quantity can be interpreted as the transition matrix element of $T^{+i}$ at the transverse position
$\bm{b}$ between baryon states localized in transverse space at the
origin \cite{Burkardt:2000za,Burkardt:2002hr,Granados:2013moa}. The different masses $m'\neq m$
do not affect the preparation of the localized states because the description of the transverse motion
in light-front quantization is independent of the mass, as in a non-relativistic system.
We define the longitudinal transition AM as (the superscript $z$ denotes the 3-component)
\begin{align}
2 S^z (\Sthreep, \Sthree ) \, 
J_{B \rightarrow B'} 
\equiv \frac{1}{2 p^+} \int d^2 b \; \left[ \bm{b} \times \bm{T}^{+T}(\bm{b}|B', B) \right]^z ,
\label{J_def}
\end{align}
where the factor $S^z (\Sthreep, \Sthree )$ accounts for the kinematic spin dependence (to be specified below)
and $J_{B \rightarrow B'}$ is independent of the spin projections $\Sthree, \Sthreep$ (reduced matrix element).
Equation~(\ref{J_def}) generalizes the light-front AM definition for diagonal transitions discussed in 
Refs.~\cite{Lorce:2017wkb,Adhikari:2016dir,Granados:2019zjw}. The integrand
can be interpreted as the transverse coordinate space density of AM and gives rise to a simple
mechanical picture \cite{Lorce:2017wkb,Adhikari:2016dir,Granados:2019zjw}.
In terms of the transverse momentum-dependent matrix element Eq.~(\ref{T_matrix_Delta}), the AM Eq.~(\ref{J_def})
is expressed as 
\begin{align}
2 S^z (\Sthreep, \Sthree ) \, 
J_{B \rightarrow B'} =
\frac{1}{2p^+}\left[ -i \frac{\partial}{\partial \bm{\Delta}_T}
\times \bm{T}^{+T}(\bm{\Delta}_T|B', B)\right]^z_{\bm{\Delta}_T = 0} .
\label{J_momentum}
\end{align}

The baryon spin states in Eq.~(\ref{T_matrix_Delta}) are chosen as light-front helicity states.
They are obtained by light-front boosts from rest-frame spin states with spins quantized in $z$-direction, and
thus effectively depend on the rest-frame spins $S, S'$ and their projections $\Sthree, \Sthreep$. The dependence
of the matrix element Eq.~(\ref{T_matrix_Delta}) on the transverse direction of $\bm{\Delta}_T$ and on the
spin projections $\Sthree$ and $\Sthreep$ is kinematic and can be made explicit by performing a
transverse multipole expansion. Showing only the dipole term (linear in $\bm{\Delta}_T$)
that gives rise to the longitudinal AM, we write
\begin{align}
\bm{T}^{+T}(\bm{\Delta}_T | \Sthreep, \Sthree) &= 2p^+ [i \bm{\Delta}_T \times \bm{e}_z
S^z (\Sthreep, \Sthree)] \; F_1 (-\bm{\Delta}_T^2) + \ldots,
\label{lf_multipole}
\end{align}
where $\bm{e}_z$ is the unit vector in the $z$-direction, $S^z(\Sthreep, \Sthree)$ is the
$z$-component of a spin 3-vector depending on the rest-frame spin projections $\Sthree$ and $\Sthreep$
(in a form that is specific to the spins $S$ and $S'$), and $F_1(t)$ is a form factor.
For a $\frac{1}{2} \rightarrow \frac{1}{2}$ transition ($N \rightarrow N$), the $z$-component of
the spin vector is 
\begin{align}
S^z (\Sthreep, \Sthree) \; &\equiv \; \Sthree \, \delta (\Sthree, \Sthreep) \; = \; \pm {\textstyle\frac{1}{2}} .
\end{align}
More generally, for any transition between states of the same spin $S \rightarrow S$
with $S = \frac{1}{2}, \frac{3}{2}, \ldots$ ($N \rightarrow N, \Delta \rightarrow \Delta, \ldots$),
the $z$-component of the spin vector is 
\begin{align}
S^z (\Sthreep, \Sthree ) \; &= \; \sqrt{S(S+1)} \; \langle S \Sthree , 1 0 | S \Sthreep \rangle ,
\label{spin_vector_same}
\end{align}
where $\langle j_1 m_1 j_2 m_2 | J M \rangle$ are the vector coupling coefficients.
For transitions between states with spins $|S' - S| = 0, 1$ such as $\frac{1}{2} \rightarrow \frac{3}{2}$
($N \rightarrow \Delta$) we define the spin vector such that
\begin{align}
S^z (\Sthreep, \Sthree ) \; &= \; \sqrt{S(S+1)} \; \sqrt{\frac{2S + 1}{2S' + 1}} \;
\langle S \Sthree, 1 0 | S' \Sthreep\rangle ,
\label{spin_vector_transition}
\end{align}
which reduces to Eq.~(\ref{spin_vector_same}) if $S' = S$
(with this definition the form
factor is independent of $S$ and $S'$ in large-$N_c$ limit; see below).
In each case, the AM obtained from Eq.~(\ref{lf_multipole}) with Eq.~(\ref{J_momentum})
is then given by the form factor at $t = 0$
\begin{align}
J_{B \rightarrow B'} &= F_1(0).
\end{align}
The normalization of $J_{B \rightarrow B'}$ adopted here is such that the spin sum rule for the nucleon,
which involves the isoscalar quark and the gluon EMT, is \cite{Lorce:2017wkb,Granados:2019zjw}
\begin{align}
J_{N \rightarrow N}^S + J_{N \rightarrow N}^{\rm glu} = {\textstyle \frac{1}{2}}.
\label{J_sumrule}
\end{align}

The isospin dependence of the matrix element Eq.~(\ref{T_matrix}) is governed by the usual selection rules.
The isoscalar component of the EMT in Eq.~(\ref{isoscalar_isovector}) connects only states with $I'=I$,
while the isovector component can connect states with $|I' - I| = 0$ or 1. In both cases the isospin
projection is conserved, $I_3' = I_3$. More generally, the matrix element of the isovector operator
in Eq.~(\ref{isoscalar_isovector}) is proportional to
$\langle I I_3, 1 0 | I' I_3 \rangle$; for the transition $\frac{1}{2} \rightarrow \frac{3}{2}$
($N \rightarrow \Delta$) this factor is
$\langle \frac{1}{2} I_3, 1 0 | \frac{3}{2} I_3 \rangle = \sqrt{2/3}$ for both $I_3 = \pm \frac{1}{2}$.
These isospin factors are included in the values of the transition AM defined in Eqs.~(\ref{J_def})
and (\ref{J_momentum}).
\section{$N \rightarrow \Delta$ transition angular momentum in $1/N_c$ expansion}
\label{sec:ndelta_ncexp}
In the $N_c \rightarrow \infty$ limit of QCD, the dynamics is characterized by the emergent SU(2$N_f$)
spin-flavor symmetry (here $N_f = 2$) \cite{Gervais:1983wq,Gervais:1984rc,Dashen:1993as,Dashen:1993jt,Dashen:1994qi}.
The $N$ and $\Delta$ baryons appear in the totally symmetric representation with spin/isospin
$S = I = \frac{1}{2}, \frac{3}{2}, ...$. Transition matrix elements of QCD operators between these states are thus
connected by the symmetry. A systematic expansion in $1/N_c$ can be performed, including subleading
corrections \cite{Dashen:1993jt,Dashen:1994qi}. The baryon masses are $m_{N, \Delta} = \mathcal{O}(N_c)$,
and the mass splitting is $m_\Delta - m_N = \mathcal{O}(N_c^{-1})$. Here we apply this method to the
transition matrix elements of the EMT and the transition AM.

The $1/N_c$ expansion of baryon transition matrix elements is performed in a class of frames where the
baryons have 3-momenta $|\bm{p}|, |\bm{p}'| = \mathcal{O}(N_c^0)$, so that the velocities are
parametrically small, $|\bm{p}|/m, |\bm{p}'|/m' = \mathcal{O}(N_c^{-1})$.
The 3-momentum transfer is $\bm{\Delta} = \bm{p}' - \bm{p} = \mathcal{O}(N_c^0)$, and the
energy transfer for transitions within the same multiplet is $\Delta^0 = E' - E = \mathcal{O}(N_c^{-1})$.
In particular, for the $1/N_c$ expansion of the EMT we choose the symmetric frame where the average
baryon 3-momentum is zero, $\bm{P} = (\bm{p}' + \bm{p})/2 = 0$ (generalized Breit frame).
In the notation $p = (p^0, \bm{p})$, the 4-momentum components are given by
\begin{align}
p &= (E, -\bm{\Delta}/2), &E = \sqrt{m^2 + |\bm{\Delta}|^2/4},
\nonumber \\
p' &= (E', \phantom{-}\bm{\Delta}/2), & E' = \sqrt{m^{\prime 2} + |\bm{\Delta}|^2/4},
\nonumber \\[1ex]
\Delta &= (E' - E, \bm{\Delta} ).
\label{frame_3d}
\end{align}
In this frame the only 3-vector arising from the particle momenta is the momentum transfer $\bm{\Delta}$.
The matrix elements of the tensor operator obey standard angular momentum selection rules, and a
multipole expansion can be performed for the components
\begin{align}
T^{00}, \; T^{0k}, \; T^{kl} \hspace{2em} (k,l = 1, 2, 3).
\end{align}

%
%
\begin{figure}[t]
\begin{center}
\includegraphics[width=.8\columnwidth]{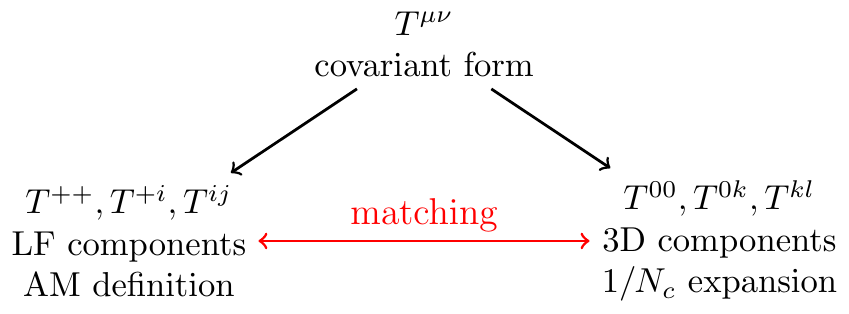}
\end{center}
\caption[]{Matching of light-front and 3D components of the EMT.}
\label{fig:matching}
\end{figure}
The $1/N_c$ expansion of the light-front components of the EMT of Sec.~\ref{sec:transition_am}
is obtained by matching the ordinary 4-vector components
with the light-front components in the same frame (see Fig.~\ref{fig:matching}).
The symmetric frame Eq.~(\ref{frame_3d}) is contained in the class of $\Delta^+ = 0$ frames Eq.~(\ref{frame_lf});
namely, it is the frame with
\begin{align}
p^+ = \sqrt{P^2} =  {\textstyle \sqrt{(m ^2 + m^{\prime 2})/2 - t/4}} .
\end{align}
The light-front energy transfer in this frame is $\Delta^- = \mathcal{O}(N_c^{-1})$, see Eq.~(\ref{frame_lf}),
and thus small and of the same order as the ordinary energy transfer $\Delta^0$.
The light-front components of the EMT are then calculated as
\begin{align}
T^{+i} &= T^{0i} + T^{3i} \hspace{1em} (i = 1, 2) \hspace{1em} \text{etc.}
\end{align}
The matching procedure performed here is unambiguous because one is dealing with on-shell matrix elements.
It automatically implements 3-dimensional rotational invariance, which is not manifest in the
light-front formulation and must be implemented through conditions on light-front matrix elements (angular conditions).
It is analogous to the procedure used in nuclear physics for matching the light-front nuclear wave function with
the 3-dimensional nonrelativistic wave function \cite{Frankfurt:1988nt,Lev:1998qz}.

We have computed the $1/N_c$ expansion of the 3-dimensional multipoles of the EMT in the symmetric
frame Eq.~(\ref{frame_3d}) using a method based on the soliton picture of large-$N_c$
baryons \cite{Goeke:2001tz,Schweitzer:2016jmd}; equivalently one can use methods based on the
algebra of the spin-flavor symmetry group \cite{Dashen:1993jt,Dashen:1994qi}. The full results will
be presented elsewhere \cite{multipole}; in the following we quote only the multipoles relevant to
the AM. In leading order of $1/N_c$, the matrix elements of the isoscalar and isovector
components [see Eq.(\ref{isoscalar_isovector})] of $T^{0k}$ are of the form
\begin{align}
&\langle B', \bm{\Delta}/2 | (\hat{T}^S)^{0k} | B, -\bm{\Delta}/2 \rangle = 2m^{2}
\langle S^i \rangle_{B'B} \left[ i\epsilon^{kil} \frac{\Delta^{l}}{m} \mathcal{J}_{1}^S(t) + ... \right] ,
\label{T0k_S}
\\
&\langle B', \bm{\Delta}/2 | (\hat{T}^V)^{0k} | B, -\bm{\Delta}/2 \rangle = 2m^{2}
\langle D^{3i} \rangle_{B'B} \left[ i\epsilon^{kil} \frac{\Delta^{l}}{m} \mathcal{J}_{1}^V(t)
+ ... \right] ,
\label{T0k_V}
\end{align}
where we have omitted spin-independent terms $\propto \Delta^k$ that do not contribute to the AM.
The spin/isospin dependence is contained in the structures (here $i = 0, \pm 1$ denote the spherical
3-vector components)
\begin{align}
\langle S^i \rangle_{B'B} &=
\sqrt{S(S+1)} \; \langle S \Sthree, 1 i | S' \Sthreep \rangle \;
\delta_{S'S} \delta_{I'I} \delta_{\Ithreep I_3} ,
\label{collective_spin}
\\
\langle D^{3i} \rangle_{B'B} &=
- \sqrt{\frac{2S + 1}{2S' + 1}} \; \langle S \Sthree, 1 i | S' \Sthreep \rangle \;
\langle I I_3, 1 0 | I' \Ithreep \rangle .
\label{collective_spin_isospin}
\end{align}
$S^i$ has only matrix elements between same spin/isospin, while $D^{3i}$ can connect states with spin/isospin
differing by one.\footnote{The matrix
elements Eq.~(\ref{collective_spin}) and (\ref{collective_spin_isospin})
appear from the collective quantization of the soliton rotations \cite{Goeke:2001tz,Schweitzer:2016jmd}.
In the formulation of the $1/N_c$ expansion based on the SU(4) spin-flavor symmetry
\cite{Dashen:1993as,Dashen:1993jt,Dashen:1994qi}, $\langle D^{ai} \rangle_{B'B} (i, a = 1, 2, 3)$
is related to the matrix element of the spin-flavor generator $G^{ia}$, namely
$\langle D^{ai} \rangle_{B'B}= -4/(N_{c}+2) \langle G^{ia} \rangle_{B'B}+ \mathcal{O}(N^{-2}_{c})$.}
Thus $(\hat{T})^S$ in Eq.~(\ref{T0k_S}) contributes only to $N \rightarrow N$ and $\Delta \rightarrow \Delta$
transitions, while $N \rightarrow \Delta$ transitions arise only from $(\hat{T})^V$ in Eq.~(\ref{T0k_V}).
$\mathcal{J}_{1}^{S, V}(t)$ in Eqs.~(\ref{T0k_S}) and (\ref{T0k_V}) are the isoscalar and isovector dipole
form factors. They are found to be of the order \cite{multipole}
\begin{align}
\mathcal{J}_{1}^S = \mathcal{O}(N_c^0),
\hspace{2em}
\mathcal{J}_{1}^V = \mathcal{O}(N_c).
\label{J_S_J_V}
\end{align}
The matrix elements of $T^{3k}$ are suppressed by $1/N_c$ compared to those of $T^{0k}$ in both the isoscalar
and isovector sector. The light-front component $T^{+i}$ is therefore given by $T^{0k}$ in leading order of the
$1/N_c$ expansion, and we can compute the AM Eq.~(\ref{J_momentum}) from Eqs.~(\ref{T0k_V})--(\ref{J_S_J_V}).
We find:

(i)~The isovector AM in the nucleon is leading in $1/N_c$; the isoscalar is subleading.
\begin{align}
J^{{S}}_{N\to N} = \mathcal{J}^{S}_{1}(0) = \mathcal{O}(N_c^0),
\hspace{1em}
J^{{V}}_{p\to p} = -\frac{2}{3}\mathcal{J}^{{V}}_{1}(0) = \mathcal{O}(N_c) .
\label{jv_nc}
\end{align}
This explains the observed large flavor asymmetry of the AM. Note that this scaling is consistent with
that of the quark spin contribution to the nucleon spin as given by the axial coupling,
$g_A^S = \mathcal{O}(N_c^0)$ and $g_A^V = \mathcal{O}(N_c^1)$.

(ii)~The isoscalar component of the AM in the nucleon and $\Delta$ are related by
\begin{align}
J^{{S}}_{N\to N} = J^{{S}}_{\Delta \to \Delta} =  \mathcal{J}^{S}_{1}(0) .
\end{align}
This provides insight into the spin structure of $\Delta$ resonance. Note that this relation
is consistent with the spin sum rule for the $\Delta$ state.

(iii) The isovector AM in the nucleon, the AM in the $N \rightarrow \Delta$ transitions, and the
isovector AM in the $\Delta$ are related by
\begin{align}
J^{V}_{p\to p} = \frac{1}{\sqrt{2}} J^{{V}}_{p\to \Delta^+}
= 5 J^{V}_{\Delta^+ \to \Delta^+} = -\frac{2}{3} \mathcal{J}^{V}_1 (0).
\end{align}
This suggests that the $N \rightarrow \Delta$ transition AM is large and provides a way to probe the
isovector nucleon AM with $N \rightarrow \Delta$ transition measurements.
\section{$N \rightarrow \Delta$ transition angular momentum from lattice QCD}
%
%
\begin{table}[t]
\setlength{\tabcolsep}{5pt}
\renewcommand{\arraystretch}{1.2}
\begin{tabular}{c|cc|ccc} 
\hline
\hline
Lattice QCD & $J^{S}_{p\to p}$ & $J^{S}_{\Delta^{+}\to \Delta^{+}}$ & $J^{V}_{p\to p}$ & $J^{V}_{p\to \Delta^{+}}$ &
$J^{V}_{\Delta^{+}\to \Delta^{+}}$   \\
\hline
  \cite{Gockeler:2003jfa}~$\mu^{2}=4 \, \mathrm{GeV}^{2}$ 
  & $0.33^\ast$ & $0.33$ & $0.41^\ast$ & $0.58$ & $0.08$ \\
  \cite{LHPC:2007blg}~$\mu^{2}=4 \, \mathrm{GeV}^{2}$
  & $0.21^\ast$ & $0.21$ & $0.22^\ast$ & $0.30$ & $0.04$ \\
  \cite{LHPC:2010jcs}~$\mu^{2}=4 \, \mathrm{GeV}^{2}$
  & $0.24^\ast$ & $0.24$ & $0.23^\ast$ & $0.33$ & $0.05$ \\
  \cite{Bali:2018zgl}~$\mu^{2}=1 \, \mathrm{GeV}^{2}$
  & $-$ & $-$ & $0.23^\ast$ & $0.33$ & $0.05$ \\
  \cite{Alexandrou:2019ali}~$\mu^{2}=4 \, \mathrm{GeV}^{2}$
  & $-$ & $-$ & $0.17^\ast$ & $0.24$ & $0.03$ \\
\hline 
\hline
\end{tabular}
\caption{Estimates of the isoscalar and the isovector AM for $p\to p$, $p\to\Delta^{+}$ and $\Delta^{+}\to\Delta^{+}$
obtained from lattice QCD data on $J^S_{p\to p}$ and $J^V_{p\to p}$ and the relations provided by the
leading-order $1/N_{c}$ expansion.
Here $S, V \equiv u \pm d$, and the nucleon matrix elements are normalized as in Eq.~(\ref{J_sumrule}).
Input values are marked by an asterisk $^\ast$.} 
\label{tab:lattice}
\end{table}
We now evaluate the transition AM using the leading-order $1/N_c$ expansion relations together
with lattice QCD results for the EMT matrix elements. This provides a numerical estimate of
the transition AM and illustrates the dominance of the isovector component of the nucleon AM.
Lattice QCD calculations of $N \rightarrow N$ matrix elements of the symmetric EMT Eq.~(\ref{T_symmetric})
have been performed in various setups (fermion implementation, normalization scale, pion mass)
\cite{Gockeler:2003jfa,LHPC:2007blg,LHPC:2010jcs,Bali:2018zgl,Alexandrou:2019ali}.
Using these as input, we obtain the values listed in Table~\ref{tab:lattice}.
One observes that a sizable isovector component of the nucleon AM is obtained in all lattice calculations
(similar large values are obtained in the chiral quark-soliton model \cite{Won:2023rec}).
Note that the lattice results for the isoscalar nucleon AM in Refs.~\cite{Gockeler:2003jfa,LHPC:2007blg,LHPC:2010jcs}
are more uncertain than the isovector, as they involve disconnected diagrams and require careful treatment
of the mixing of quark and gluon operators. Furthermore, when comparing the $1/N_c$ expansion with numerical
values of matrix elements, one needs to keep in mind that it is a parametric expansion, and that the numerical
values are determined not only by the power of $1/N_c$ but also coefficients of order unity.
\section{$N \rightarrow \Delta$ transition angular momentum from GPDs}
We now connect the $N \rightarrow \Delta$ transition AM with the transition GPDs measured in hard
exclusive electroproduction processes such as DVCS $eN \rightarrow e'\gamma\Delta$ \cite{Semenov-Tian-Shansky:2023bsy}.
This opens the prospect of future experimental studies of the transition AM.
QCD factorization at leading-twist accuracy expresses the amplitudes of hard exclusive processes
in terms of matrix elements of quark light-ray (or partonic)
operators of the type \cite{Goeke:2001tz,Diehl:2003ny,Belitsky:2005qn}
\begin{align}
\hat O_f (z) &= \bar{\psi}_f(-z/2) [-z/2, z/2] \slashed{z} \psi_f(z/2) ,
\label{partonic_operator}
\end{align}
where $z$ is a light-like 4-vector ($z^2 = 0$) and $[-z/2, z/2]$ denotes the gauge link operator.
The non-local operator Eq.~(\ref{partonic_operator}) can be represented as a power series in the
distance $z$, 
\begin{align}
\hat O_f (z) &= z^\alpha \bar{\psi}_f (0) \gamma_\alpha \psi_f (0) +
z^\alpha z^\beta  \bar\psi_f (0) \gamma_{\{\alpha} \overleftrightarrow{\nabla}_{\beta\}} \psi_f (0)
+ \ldots,
\label{partonic_expansion}
\end{align}
where the coefficients are local operators representing totally symmetric traceless tensors of
spin $n \geq 1$ (twist-2 operators).
The spin-2 operator coincides with the symmetric EMT Eq.~(\ref{T_symmetric}). The light-like vector $z$
is chosen such that it has light-front components along the ``minus'' direction, $z^- \neq 0, z^+ = \bm{z}_T = 0$.
The expansion Eq.~(\ref{partonic_expansion}) thus involves the light-front component $T^{++}$ of the EMT.
In our approach based on the $1/N_c$ expansion, this light-front component can be related to the
3D components $T^{00}, T^{0i}, T^{ij}$, and in this way be connected with the light-front component
$T^{+i}$ entering in the transition AM Eq.~(\ref{J_def}). This establishes a connection between the
transition AM defined by Eq.~(\ref{J_def}) and the leading-twist partonic operators Eq.~(\ref{partonic_operator}).

The transition matrix element of the isovector light-ray operator Eq.~(\ref{partonic_operator})
between $N$ and $\Delta$ states (here, between $p$ and $\Delta^+$ states) is parametrized covariantly
through the spectral representation
\begin{align}
\langle \Delta^+, p' | \hat{O}^V(z) | p, p \rangle
& = {\textstyle \sqrt{\frac{2}{3}}} \sum_{I=M,E,C} \; \int_{-1}^1 dx \, e^{-i x P\cdot z} \;
H_{I}(x,\xi,t)
\nonumber \\
& \times 
\overline{u}^{\alpha}(p', \Sthreep) \, (\mathcal{K}_{I})_{\alpha\beta} \, z^{\beta} \, {u}(p,S_{3}) .
\label{eq:matel1}
\end{align}
$u^\alpha$ is the spin-$\frac{3}{2}$ Rarita-Schwinger vector-bispinor of the $\Delta$, and
$u$ is the spin-$\frac{1}{2}$ bispinor of the nucleon. For the invariant bilinear forms in
the decomposition in Eq.~(\ref{eq:matel1}), various choices are possible (see also discussion below).
Here we use the tensors as defined in Ref.~\cite{Goeke:2001tz}. The magnetic tensor ($M$)
is
\begin{align}
(\mathcal{K}_{M})^{\alpha\beta} &=
\frac{3(m_{\Delta}+m_{N})}{2m_{N}[(m_{\Delta}+m_{N})^{2}-t]} \; i
\varepsilon^{\alpha\beta\gamma\delta} P_{\gamma} \Delta_{\delta} ;
\end{align}
the other structures are given in Ref.~\cite{Goeke:2001tz}. The GPDs $H_{I}(x, \xi, t)$ in Eq.~(\ref{eq:matel1})
depend on the spectral variable $x$, the light-cone momentum transfer
$\xi \equiv - \Delta \cdot z / (2 P \cdot z)$, and the invariant momentum transfer $t$.
They are defined such that their first moments satisfy the relations (sum rules)
\begin{align}
\int^{1}_{-1} dx \, H_{M,E,C}(x,\xi,t) &= 2G^{*}_{M,E,C}(t),
\label{H_M_sumrule}
\end{align}
where $G^{*}_{M, E, C}(t)$ are the $\gamma N \Delta$ transition form factors of Ref.~\cite{Jones:1972ky},
defined by multipole expansion of the decay $\Delta \rightarrow \gamma N$ in the $\Delta$ rest frame
(magnetic dipole, electric quadrupole, and Coloumb quadrupole form factors).

In the context of the $1/N_c$ expansion we can now relate the $N$--$\Delta$ transition AM
of Sec.~\ref{sec:transition_am} to the second moments of the GPDs of Eq.~(\ref{eq:matel1}).
The $1/N_c$ expansion of the GPDs is performed in the parametric regime where $x, \xi = \mathcal{O}(N_c^{-1})$
and $t = \mathcal{O}(N_c^0)$ \cite{Diakonov:1996sr,Petrov:1998kf} and can be implemented using the
techniques described in Refs.~\cite{Goeke:2001tz,Schweitzer:2016jmd}. The dominant $N \rightarrow \Delta$
GPD is the magnetic GPD $H_M$. In the large-$N_c$ limit it scales as \cite{Goeke:2001tz}
\begin{align}
H_{M}(x,\xi,t) &\sim N_c^3 \times \mathrm{function}(N_{c}x,N_{c}\xi,t).
\end{align}
The power $N_c^3$ multiplying the scaling function can be inferred from the known $N_c$ scaling of
the $N$--$\Delta$ transition magnetic moment, which determines the first moment of $H_M$ through
Eq.~(\ref{H_M_sumrule}),
\begin{align}
\frac{\mu_{\Delta N}}{m} \equiv \frac{G_M^\ast(0)}{m} = \frac{1}{2 m}
\int^{1}_{-1} dx \, H_M(x,\xi, 0) = \mathcal{O}(N_c),
\end{align}
where $m = \mathcal{O}(N_c)$ is the common baryon mass in the large-$N_c$ limit.
In leading order of $1/N_c$ we obtain
\begin{align}
\int^{1}_{-1} dx \, x H_{M}(x,\xi,0) &=
2 J^{V}_{p\to \Delta^{+}} =
-\frac{4\sqrt{2}}{3} \mathcal{J}^{V}_{1}(0),
\label{H_M_AM}
\end{align}
which agrees with the $N_c$ scaling established earlier, Eq.~(\ref{jv_nc}).\footnote{The coefficient
in Eq.~(\ref{H_M_AM}) agrees with the one in the large-$N_c$ relation between the $N \rightarrow \Delta$
and $N \rightarrow N$ GPDs quoted in Ref.~\cite{Pascalutsa:2006ne}, but disagrees with the one quoted in
Ref.~\cite{Goeke:2001tz}.} The derivation uses
the covariant decomposition of the transition matrix elements of the EMT of Ref.~\cite{Kim:2022bwn}.
This provides the desired connection between the $N \rightarrow \Delta$ transition AM as defined
in Eq.~(\ref{J_def}) and the second moment of the transition GPDs.

In this study we have defined the transition AM through the $T^{+i}$ component of the EMT, which
can be interpreted as the cross product of momentum and distance in the transverse plane.
The AM can be defined alternatively through the $T^{++}$ component of the EMT, which can be understood
as a dipole distortion of the two-dimensional momentum distribution when the baryon spins are polarized
in the transverse direction. This definition of the transition AM will be explored
elsewhere \cite{multipole}.
\section{Discussion}
In this work we have introduced the concept of transition AM and applied it to $N\rightarrow \Delta$
transitions in the context of the $1/N_c$ expansion. We want to discuss the significance and
limitations of the present results and possible future extensions.

The present calculations are limited to the leading order of the $1/N_c$ expansion. At this level the
$N$--$\Delta$ mass difference can be neglected, and the relation between the light-front components and the
3-dimensional multipoles of the EMT involve only a single structure. However, the method developed in
Sec.~\ref{sec:ndelta_ncexp} is general and permits also the calculation of subleading terms. They include
``dynamical'' corrections due to $1/N_c$ suppressed structures, and ``kinematic'' corrections
due to the baryon motion and finite masses. Computing these corrections will be the objective of
future work.

The present study uses the symmetric version of the EMT, which measures the total AM of field
configurations in QCD \cite{Lorce:2017wkb}. Separation of orbital and spin AM in $B \rightarrow B'$
transitions would be possible by extending the definitions Eq.~(\ref{J_def}) et seq.\ to the
non-symmetric EMT and the spin operator \cite{Lorce:2017wkb}. Our results show that the total AM
and the quark spin (represented by the axial current) have the same $1/N_c$ scaling in the isoscalar
and isovector sector, see Eq.~(\ref{jv_nc}), which is natural and required for consistent scaling
of the isoscalar spin sum rule.
The separation of spin and orbital AM is a question of dynamics and can be studied with
dynamical models that are consistent with the $1/N_c$ expansion, such as the chiral quark-soliton model.

The operators describing the quark contributions to the EMT and the AM contributions derived from it
are scale dependent; only the sum of isoscalar quark and gluon AM is protected by the spin sum rule
and scale-independent; see e.g.\ Refs.~\cite{Ji:1995cu,Thomas:2008ga}. An advantage of the isovector
component is that the scale dependence is much weaker than that of the isoscalar (or of individual quark
flavor components), as there is no mixing with gluon operators. The scale dependence can be taken
into account in a more quantitative analysis.

Some comments are in order regarding the conservation of the EMT. Only the total EMT of QCD, given by the
sum of the isoscalar quark and gluon tensors, is a conserved current as obtained from Noether's theorem.
The isovector quark part studied here is generally not conserved. This circumstance needs to be taken
into account when performing a covariant decomposition of the transition matrix elements of the EMT.
In the present study we work directly with the light-front components and 3D multipoles, where this problem
does not arise. But one should be aware of it when comparing with the formulation in terms of the
covariant decomposition with invariant form factors.

The definition of the $N \rightarrow \Delta$ transition GPDs of Ref.~\cite{Goeke:2001tz} and other works
refers to the $\gamma N \Delta$ transition form factors of Ref.~\cite{Jones:1972ky}, which are defined
through a multipole expansion of the decay $\Delta \rightarrow \gamma N$ in the $\Delta$ rest frame.
While this frame can be used in the entire physical region of $t < (m_\Delta - m_N)^2,$ it does not
appear natural for the definition of light-front transition matrix elements. It would be worth to revisit the
definition of the transition GPDs using the class of frames introduced in Sec.~\ref{sec:transition_am}.

In this work we have described a method for computing the $1/N_c$ expansion of light-front tensor operators
by matching the light-front components with 3D components in a special frame. The procedure implements
3-dimensional rotational invariance of the light-front components (which is encoded in the matrix elements of
the 3D components) order-by-order in $1/N_c$. The method is general and can be extended to other states and
operators than those considered here. It can be applied to matrix elements of the EMT in hadronic states
with higher spin. A particular advantage here is that it does not require the covariant decomposition
of the matrix element in terms of invariant form factors, which becomes very cumbersome for higher spins.
The method can also be applied to other tensor operators, such as the generalized form factors
of twist-2 spin-$n$ operators.

This material is based upon work supported by the U.S.~Department of Energy, Office of Science,
Office of Nuclear Physics under contract DE-AC05-06OR23177, and by the National Science Foundation,
Grant Number PHY 1913562. This work is also supported by the Basic Science Research Program through
the National Research Foundation of Korea funded by the Korean government (Ministry of Education,
Science and Technology, MEST), Grant-No. 2021R1A2C2093368 and 2018R1A5A1025563.

\bibliography{am_ndelta}
\end{document}